\documentclass[fleqn,10pt]{olplainarticle}

\usepackage{url}

\title{A meso-scale cartography of the AI ecosystem}

\author[1]{Floriana Gargiulo}
\author[2]{Sylvain Fontaine}
\author[3]{Michel Dubois}
\author[4]{Paola Tubaro}
\affil[1]{CNRS-GEMASS, 59 rue Pouchet 75017, Paris}
\affil[2]{CNRS-GEMASS, 59 rue Pouchet 75017, Paris}
\affil[1]{CNRS-GEMASS, 59 rue Pouchet 75017, Paris}
\affil[2]{CREST, CNRS, IP PARIS, 91120 Palaiseau}


\begin{abstract}
In recent decades the set of knowledge, tools and practices, collectively referred to as ``artificial intelligence" (AI)', have become a mainstay of scientific research. Artificial intelligence techniques have not only developed enormously within their native areas of development (computer science, mathematics and statistics) but have also spread fast, in terms of application, to multiple areas of science and technology. In this paper we conduct a large scale analysis of artificial intelligence in science. The first question we address is the composition of what is commonly labeled AI, and how the various elements belonging to this domain are linked together. We reconstruct the internal structure of the AI ecosystem through the co-occurrence network of AI terms in publications’ abstracts and title, and we propose to distinguish between 15 different specialities of AI, with different temporal patterns. Further, we investigate the spreading of AI outside its native disciplines. We reconstruct the temporal dynamics of the diffusion of AI production in the whole scientific ecosystem and we describe the disciplinary landscape of AI applications. Finally we take a further step analyzing the role of collaborations for the interdisciplinary spreading of AI techniques. While the study of science frequently emphasizes the openness of scientific communities, we show that there are rarely any collaborations between those scholars who primarily develop AI, and those who apply it. Only a small group of researchers is able to gradually establish a bridge between these communities.  

\end{abstract}

\begin{document}

\flushbottom
\maketitle
\thispagestyle{empty}

\section*{Introduction}
Artificial intelligence (AI) is increasingly recognized as a vector of technological and scientific innovation (\cite{cockburn2018impact, bianchini2022artificial}) with a potentially strong impact on economic growth (\cite{aghion2018artificial}). A \textit{Nature} editorial (\cite{nature_2019}) describes it as one of the scientific events that shaped the last decade: ``Few fields are untouched by the machine-learning revolution, from materials science to drug exploration; quantum physics to medicine." 

The latest developments of AI, mostly as a result of the rise of Deep Learning (DL), provide indeed a unique potential to extract information from the unprecedented sources of data currently largely available in almost all scientific and technological domains. 
AI has been described as enabling a general paradigm shift toward a data immersive science (\cite{king2009automation}, \cite{kitchin2014bigdata}), based on smart machines able to grasp the hidden patterns and relationships from large masses of data.  

The origins of AI are usually traced back to a renowned workshop held in 1956 in Dartmouth, where a group of scientists first used this term to define their research activities and identify a distinct research area. 
According to the definition given in 2004 by John Mc Carty, promoter of the Dartmouth workshop, AI ``is the science and engineering of making intelligent machines, especially intelligent computer programs. It is related to the similar task of using computers to understand human intelligence, but AI does not have to confine itself to methods that are biologically observable." (\cite{mccarthy2004artificial})

This top-down definition, as well as similar ones 
(\cite{annoni2018artificial,oecd,wipo2019wipo}), emphasizes the overall goals of AI but leaves open the actual meaning of “intelligence”, the scope of the AI domain, and the relationship between AI and the existing structure of scientific knowledge.

While the first question related to the definition of intelligence is highly debated and controversial in the context of AI epistemology (\cite{mccarthy1981epistemological}), our study sheds light on the second aspect and aims to reconstruct the structure of the AI research area through a bottom-up approach based on a cartography of AI--related scientific publications.

This bottom-up approach addresses two research questions: 
\begin{enumerate}
    \item what is the structure of AI as a research area, i.e. the different specialities within AI and their development over time?
    \item how has AI knowledge been dynamically embedded in traditional scientific fields?
\end{enumerate}

Previous works provide partial answers to these questions. For example (\cite{frank2019evolution}) builds a classification based on citation networks among the list of AI subfields, defined through the categories adopted in Microsoft Academic Graph (MAG). However, this categorization does not allow a deep understanding of which AI terms are explicitly present in a subdomain. (\cite{bianchini2020deep}) provide a mapping of the spreading of DL practices in science and describe in great detail the geographical and disciplinary spreading of DL, but do not address the connections of DL with other AI practices. 
Other studies did so but only from the viewpoint of specific disciplines (\cite{baum2021artificial}).

We start by establishing the semantic diversity of AI, building up a large list of terms that are related to it without distinction of context, generality, temporality, or other criteria. 
Assemblng a suitable set of keywords for a bibliometric search is in itself a complex task, simplified in our case by the possibility to rely on the multiple glossaries of AI accessible online, first of all the Wikipedia AI glossary which also contains synonyms for several terms.
We therefore build the list of AI keywords by mining a large number of AI glossaries available on the web, that represent how different actors, dealing with AI, draw the perimeter of AI.
These terms represent the semantic building blocks of AI. 

Whether we describe it as a body of knowledge, practice or tools, AI is a dynamic phenomenon that has experienced several phases in its evolution over time.
As scientific innovation in general (\cite{uzzi2013atypical}) can be viewed as a cumulative process where novelty arises from the recombination of existing building blocks, a dynamical definition of AI can be seen as the result of the recombination of its building blocks, i.e. the formation of its specialities through the recombination of AI basic terms.  
Interdisciplinary exchanges also play a central role in scientific innovation, proposing new possible building blocks and thereby opening the “adjacent possible” of scientific discoveries (\cite{kauffman2000investigations,monechi2017waves}). 
Likewise, extensions of AI arise from the recombination of pre--existing applications and by interactions with other research areas: consider the example of DL, resulting from AI research on Artificial Neural Networks recombined with the connectionist approach in cognitive science.
For this reason, understanding the embedding of AI in the scientific ecosystem provides fundamental information to grasp its building process.

In the last decade, a relevant increase of the application of AI techniques in several and diverse scientific domains has been observed, above all in relation with the development of DL (\cite{bianchini2020deep,bianchini2022artificial}). 
The common idea behind this phenomenon is that AI is spreading from its ``native" disciplines (mainly computer science, mathematics and statistics), where its key tools were designed, to a series of applications in various fields of knowledge. 
This distinction between native disciplines and application disciplines can be found in several studies (\cite{cockburn2018impact, bianchini2022artificial}).  

In this study we analyze a large corpus of papers from 1970 until 2017, extracted from Microsoft Academic Graph, using AI keywords cited by the authors and different relational structures among the scientometric data (keyword co-occurrence network, authors' collaboration network).
To characterize the keywords used in corpus selection, we define their hierarchical structure in order to distinguish the core AI terms from the most peripheral ones (mainly specific algorithms and techniques). 
We first focus on the definition of the meso--scale structure of AI, namely on the identification of the specialities of AI, their interactions and their temporal patterns (Section \ref{specialities}). 
Second, we analyze how AI is globally spreading in muliple research areas or disciplines.
A first phase of concentration of AI in the ``native'' disciplines of computer science, mathematics, and statistics can be observed at the end of the 1980s, after the so-called ``AI Winter", with the emergence of expert systems and the decline of symbolic AI. 
These disciplines will remain responsible for the production of AI literature until today. However, a spreading phase started in the last decade, corresponding with the development of DL, where AI knowledge started to be largely applied to several other disciplines (Section \ref{sapplications}). 
We also show the disciplinary patterns associated with the different specialities of AI. We can observe for example that only a few specialities (like dimensionality reduction techniques and DL) were able to reach a high degree of diversity in the application ecosystem.
Finally, we highlight the collaboration mechanisms responsible for knowledge transfer from the originating domains to applications. We notice indeed that very few collaborations exist between researchers in disciplines that create AI and researchers in disciplines that only (or mainly) apply AI. The transfer of AI knowledge is largely ascribable to a core of multidisciplinary researchers mutually interacting both with AI developers and with researchers in applied disciplines (Section \ref{authors}).

\section{Data and Methods}

\subsection{AI terms}
There are several definitions of AI, and each of them implies a different perimeter of the terms or lexical units associated with it. With the effort of defining the perimeter of AI, several and diverse actors involved in its production made available online glossaries containing lists of associated keywords, with the objective of identifying the variety of terms that it covers. In particular, Wikipedia has a large list of its pages connected to AI, including synonyms.

We started by extracting the content of  the Wikipedia index page\footnote{\url{https://en.wikipedia.org/wiki/Glossary_of_artificial_intelligence}} and, after that, we performed a Google query searching for ``AI glossary”,``AI keywords”, ``AI terms”, `` AI concepts”.
We obtained a set of more than 20 specific glossaries, for example \footnote{\url{https://developers.google.com/machine-learning/glossary}}
\footnote{\url{https://machinelearning.wtf/}}
\footnote{\url{https://link.springer.com/content/pdf/bbm:978-3-319-94878-2/1.pdf}}. 

We built our original list of terms from all the keywords from these web resources, removing duplicates and lemmatizing words. 
We manually cleaned the list of keywords, removing very general words not strictly related to AI (like “software”, “algorithm”, and “self-management”). The final list includes 594 terms, mostly bigrams or trigrams, with different levels of generality. There are general terms like “machine learning” and specific algorithmic procedures like “word2vec”. The full list of terms is reported in the appendix.

\subsection{The bibliometric dataset}
The bibliometric dataset on which this article is based starts from a recent data dump of the Microsoft Academic Graph (MAG), disambiguated, and made available by M. Färber on the Zenodo platform \footnote{\url{https://zenodo.org/record/4617285#.YlUoaS2ubs0}} (\cite{farber2019microsoft}). 
From this dataset, we first selected all the papers including any of the previously identified 594 AI-terms in their abstracts or title (2,737,813 papers with associated metadata). 
From this set we only keep the papers published after 1970. This choice could appear too strict, missing almost two decades of early AI research, but it avoids the heterogeneities that would result from differences in editorial policies and scientific infrastructure in that period compared to today, notably in terms of peer reviewing. 
Additionally, we retain only studies published in or before 2017 because of a possible bias in the MAG database for later entries, which can be guessed from an unmotivated decrease in the total number of papers. 
We further filtered this dataset to the papers published in  journals or conferences indexed in the Web of Science (WoS), getting a final set of metadata for 1,090,138 papers. 
We associate to each of these papers two supplementary attributes with respect to the MAG metadata: the disciplinary fields, according to the first label (which is indeed the more specific) in the WoS classification of journals and conferences, and the list of AI keywords contained in their abstracts. 
To build the authors’ collaboration network we used the disambiguated authors’ identifiers provided by Färber et al. in the last version of the MAG database (\cite{farber2022microsoft}).

\begin{figure}[ht]
\centering
\includegraphics[width=1\linewidth]{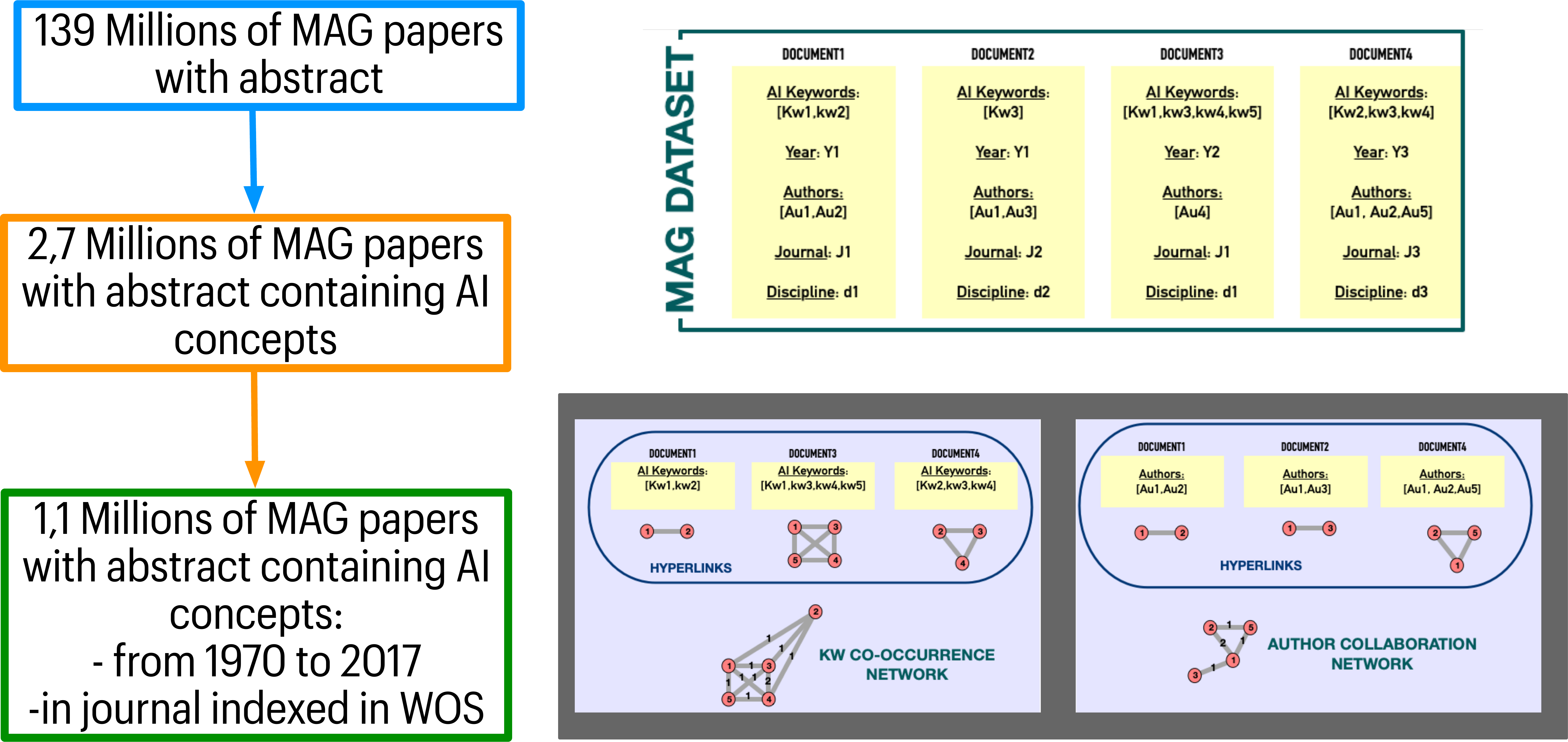}
\caption{The dataset. Left plot: filtering process of the MAG corpus. Top right plot: Structure of the AI corpus. Bottom right plot: Building process of the keywords co-occurrence network and of the author collaboration network}
\label{fig0}
\end{figure}

To summarize, the bibliometric corpus that we have constituted starting from the MAG dataset is a collection of documents having the following attributes (Figure \ref{fig0}): The list of the AI keywords contained in the abstract and/or title, the publication year, the list of authors, the journal (or conference), and the disciplinary field derived from the journal's classification and categorical structure in the WoS. 

\subsection{The network structures}
With these data, we reconstruct two different network structures: the keyword co-occurrence network (KCON) and the author collaboration network (ACN). 

These networks are directly built from the documents of the corpus as described in Figure \ref{fig0}. For the keyword co-occurrence, each document that contains more than a keyword represents an hyperedge of the graph, namely a fully connected subgraph among all the present keywords. 
These hyperedges are merged together and each link gets a weight corresponding to the number of hyperedges where it is present. 
The procedure is identical for the authors’ collaborations. Both these graphs are undirected weighted structures. 

The keyword co-occurrence graph, KCON, has 535 nodes and 24,358 edges (some less frequently used AI terms were indeed disconnected from the larger component and were therefore omitted in the rest of the analyses).
Being this network extremely dense (density=0.17) and being the weights very heterogeneous, we first apply a disparity filter on the original graph (\cite{serrano2009extracting}) to get the relevant connections and simplify the partitioning of the structure. The filtered graph DKCON has 3,276 edges. 
The author collaboration network (ACN) has 103,175 nodes and 453,137 edges. 

\subsection{The disciplinary distance matrix}
To reconstruct a distance matrix between all the WoS disciplines, we started again from the whole MAG dataset, filtered on the WoS journals. 
To reduce computation time, which would be very significant if we analyzed a single large snapshot from this extremely large dataset, we followed a procedure of producing several independent samples. Specifically, we extracted 10,000 random samples of 100,000 papers. For each paper in each random sample we extracted the list of all the referenced papers and, from the latter, the set of unique WoS disciplines relative to the references. 
From these lists of disciplines, following the same procedure adopted for the KCON and the ACN networks, we build the co-citation structure of disciplines in the sample.

Since the weighting structure of this graph (the number of co-occurrences $w_{ij}$)  is strongly related to the relative frequency of each discipline we implement a similarity measure based on the pointwise mutual information between the nodes (disciplines):
\begin{equation}
    \text{pmi}_{ij}=\text{max}\left(2\text{log}_2\left(\frac{w_{ij}}{\sum_k(w_{ik})\sum_k(w_{jk})}\right),0\right)
\end{equation}
The pointwise mutual information ranges indeed between -1 and 1; in our case, the negative values, representing a very uncorrelated situation are put to zero to obtain an indicator ranging from 0 to 1. From this similarity measure we simply obtain a distance matrix whose values are given by: $D_{ij}=1-pmi_{ij}$.

We repeat this computation for all the 10,000 samples and the final distance matrix is derived from the average values on all the samples.

\subsection{The spreading indicators}
We measure how a corpus is concentrated around the so-called native disciplines of AI (as above, computer science, mathematics and statistics) with a measure inspired from solid body mechanics, the moment of inertia: 
\begin{equation}
    m_I=\sum_i \frac{n_i}{N_{tot}}\text{min}(D_{i,CS},D_{i,Math},D_{i,Stat})^2
\end{equation}
where $i$ covers all the disciplines present in the corpus.
If the moment of inertia is small, the corpus is very concentrated around the native disciplines; if not, it is largely diffused in the disciplinary ecosystem.

To measure how AI is represented in a discipline we compare the number $n_i^{AI}$ of AI papers produced in this discipline with an expected value given by the share of publications in the given discipline ($s_i=N_{tot}^i/N_{tot}$, extracted from the whole MAG corpus) multiplied by the total number of AI publications. We define therefore the AI score of a discipline:

\begin{equation}
    z_i=\frac{n_i^{AI}-N_{tot}^{AI}s_i}{n_i^{AI}+N_{tot}^{AI}s_i}
\end{equation}
This measure ranges between -1 and 1. Positive high values of this indicator indicate that AI is more represented in the discipline than in a case in which diffusion followed a random process, and \textit{vice-versa}.
The same measure also applies at the level of journals. 

Finally, we compute for each author in the corpus an AI score, $A_I$, given by the fraction of papers published by the author in the native disciplines of computer science, mathematics and statistics.

\section{Results}

\subsection{The specialities of AI}
\label{specialities}
AI is an umbrella term encompassing a broad set of knowledge, tools and practices aimed to the general purpose of making intelligent machines and computer programs. For a comprehensive understanding of AI, it is essential to describe its specialities or thematic diversity. To do this, we adopt a bottom-up approach based on the analysis of AI related scientific publications. In particular, we study the co-occurrence of the AI keywords in the abstracts, DKCON, as described in the methods section.

\begin{figure}[ht]
\centering
\includegraphics[width=0.7\linewidth]{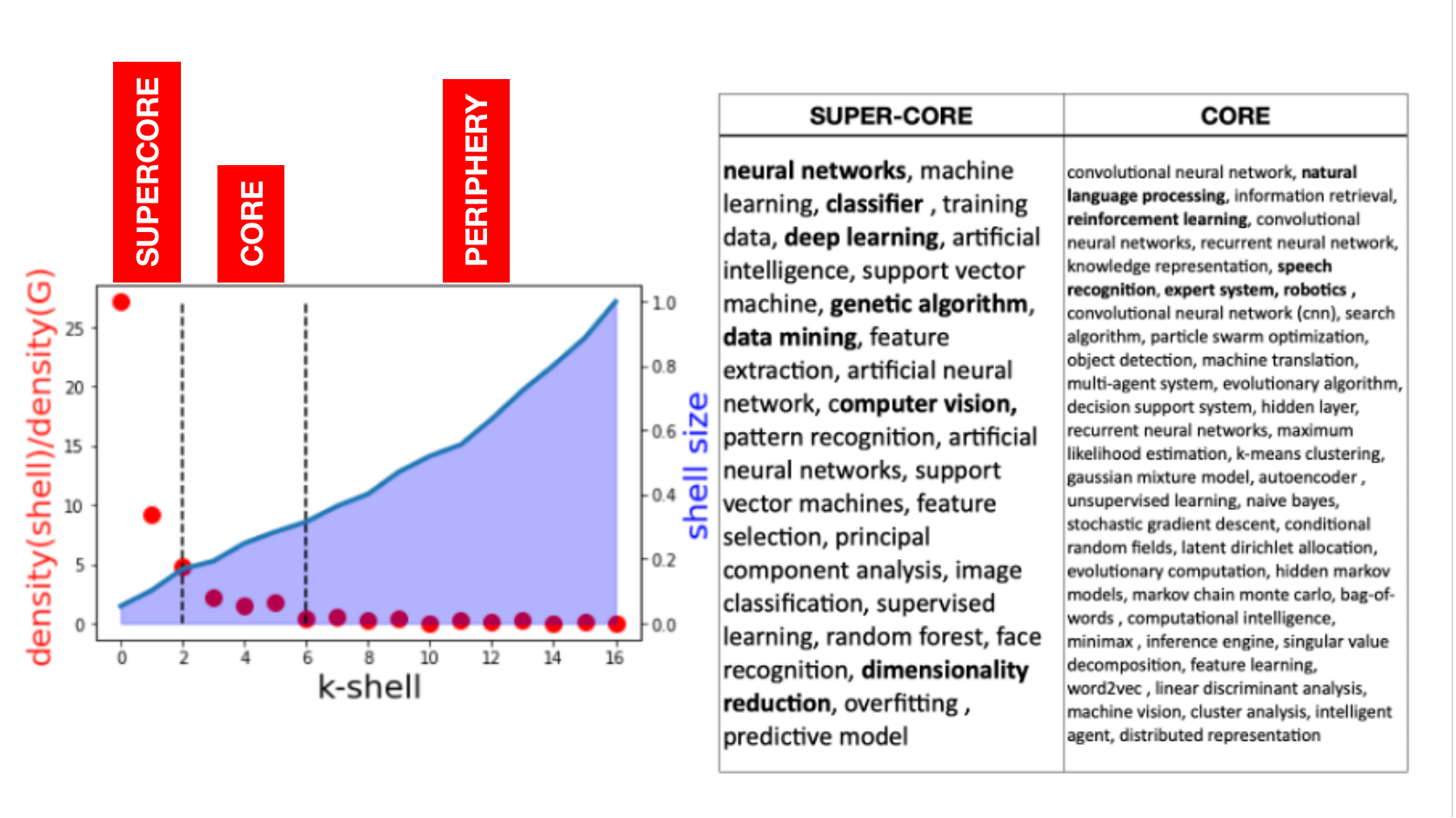}
\caption{AI term hierarchy.}
\label{fig1a}
\end{figure}

As we pointed out in the data presentation, the keywords used in the query have different levels of “generality”. We first use the filtered keyword co-occurence graph (DKCON) to identify the hierarchy of dependencies between keywords. We build the $k$-shell structure of the graph and we calculate the internal density of each shell, compared to the density of the whole DKCON graph. This analysis allows to distinguish three dimensions: the super core, the core and the periphery. 
Figure \ref{fig1a} shows that the first two shells are very dense: they include a group of 25 keywords largely used and tightly connected among them. We call these first two shells the “super-core”. This class contains general AI categories (``artificial intelligence",``‘machine learning", ``DL", ``neural networks") and very popular classes of methods (``random forest", ``support vector machine").  The internal density decreases suddenly starting from the third shell and goes to zero in the most external shells, starting from the seventh one. We define shells 3-6 the ``core" and the last ones as the ``periphery". The core also contains general methods (such as ``cluster analysis", ``particle swarm optimization`", ``stochastic gradient descent") but less connected among them and hierarchically depending on super-core terms (namely connected to the corpus only through super-core terms). The periphery mostly contains specific algorithms and specific methods not connected among them but just to the more central cores.

\begin{figure}[ht]
\centering
\includegraphics[width=1\linewidth]{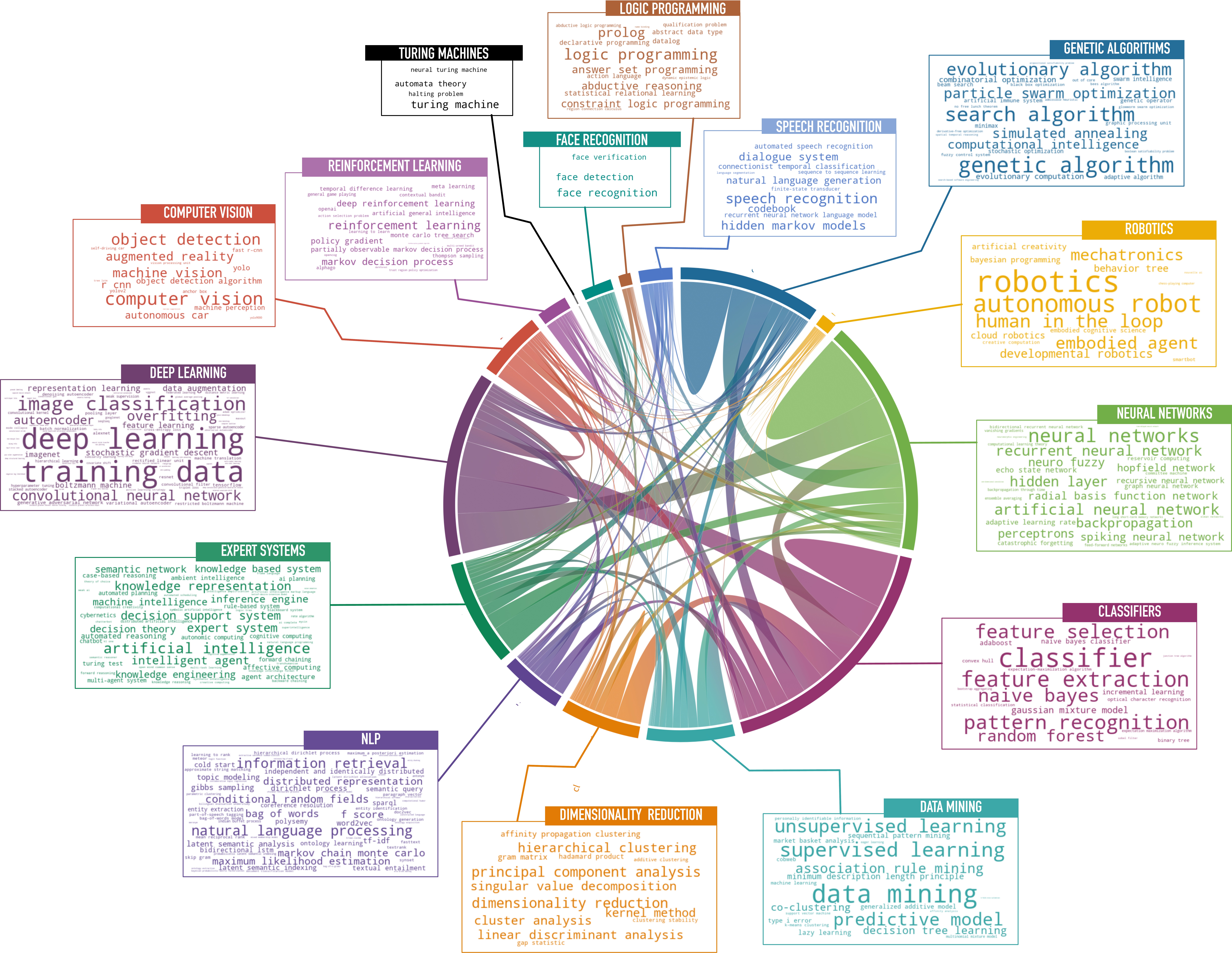}
\caption{AI specialities and their relationships.}
\label{fig1}
\end{figure}

We apply to the DKCON graph the well-known Louvain community detection algorithm (\cite{blondel2008fast}) and identify the presence of 15 meso-scale structures that correspond to a partitioning of the network at the level of specialities (Figure \ref{fig1}): 
expert systems, natural language processing, dimensionality reduction, data mining, classifiers, neural networks, robotics, genetic algorithms, speech recognition, logic programming, face recognition, Turing machines, reinforcement learning, computer vision and DL. These structures are labeled according to their internal concept characterized by the term with the highest core position.

Some of these specialities show a significant degree of openness, demonstrating a flow of knowledge from one domain to the other. For example the DL speciality has several semantic relationships both with neural networks and with classifiers. Classifiers and dimensionality reduction are strictly related, being indeed the latter a particular form of classification. Computer vision has connections with the classifier class but also with DL, as indeed computer vision was one of the first application domains of DL techniques. 

\begin{figure}[ht]
\centering
\includegraphics[width=1\linewidth]{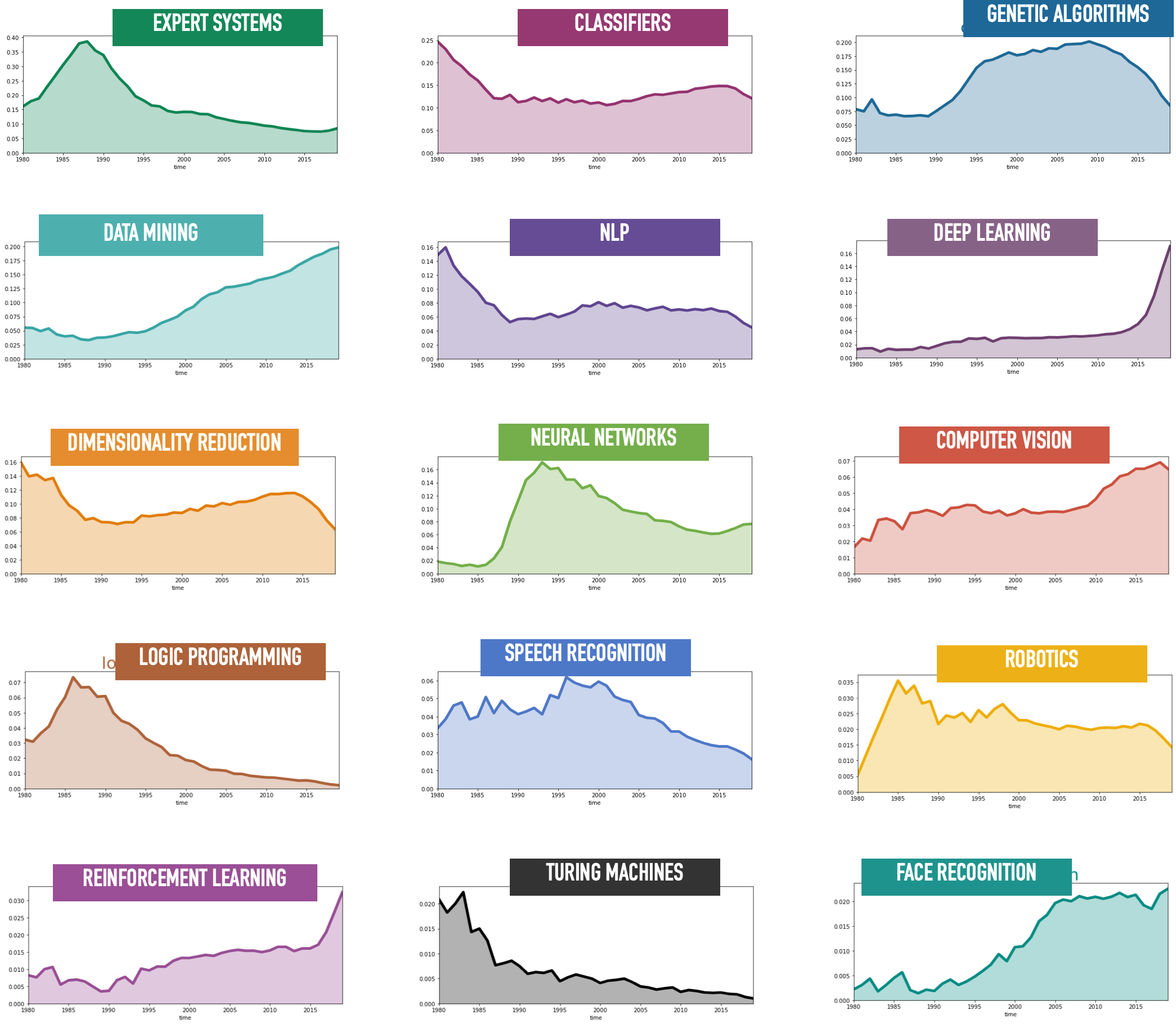}
\caption{The timeline of AI specialities.}
\label{fig2}
\end{figure}

The different AI specialities are also characterized by different temporal patterns that better define the temporality of the knowledge flows between them. In Figure \ref{fig2} we show the annual time series of the number of publications in each area. AI general terms, as well as Turing machine and logic programming (symbolic AI), widely diffused in the early days, disappeared after the AI winter -- around 1995 -- while on the contrary the “expert systems” speciality (together with agent-based systems) started to emerge. 
Just afterwards, we observe the rapid growth of specialities like neural networks, data mining, optimization and face recognition. Finally, the last two decades see a fast decrease of specialities like optimization and dimensionality reduction, parallel to the extremely fast development of DL. Analyzing the relationships among specialities, while optimization research does not seem to enter new combinations with keywords in emerging areas (indicating a gradual fading of research interest in this domain), dimensionality reduction is being gradually recombined with DL concepts.

\subsection{AI: from development to applications}
\label{sapplications}
The study of our database highlights an important, and perhaps somewhat surprising, fact : almost half of the publications included (48\%) are associated with disciplines outside the native computer science, mathematics or statistics. 
This section strictly focuses on this subset of the corpus: the applied side of AI.

\begin{figure}[ht]
\centering
\includegraphics[width=0.7\linewidth]{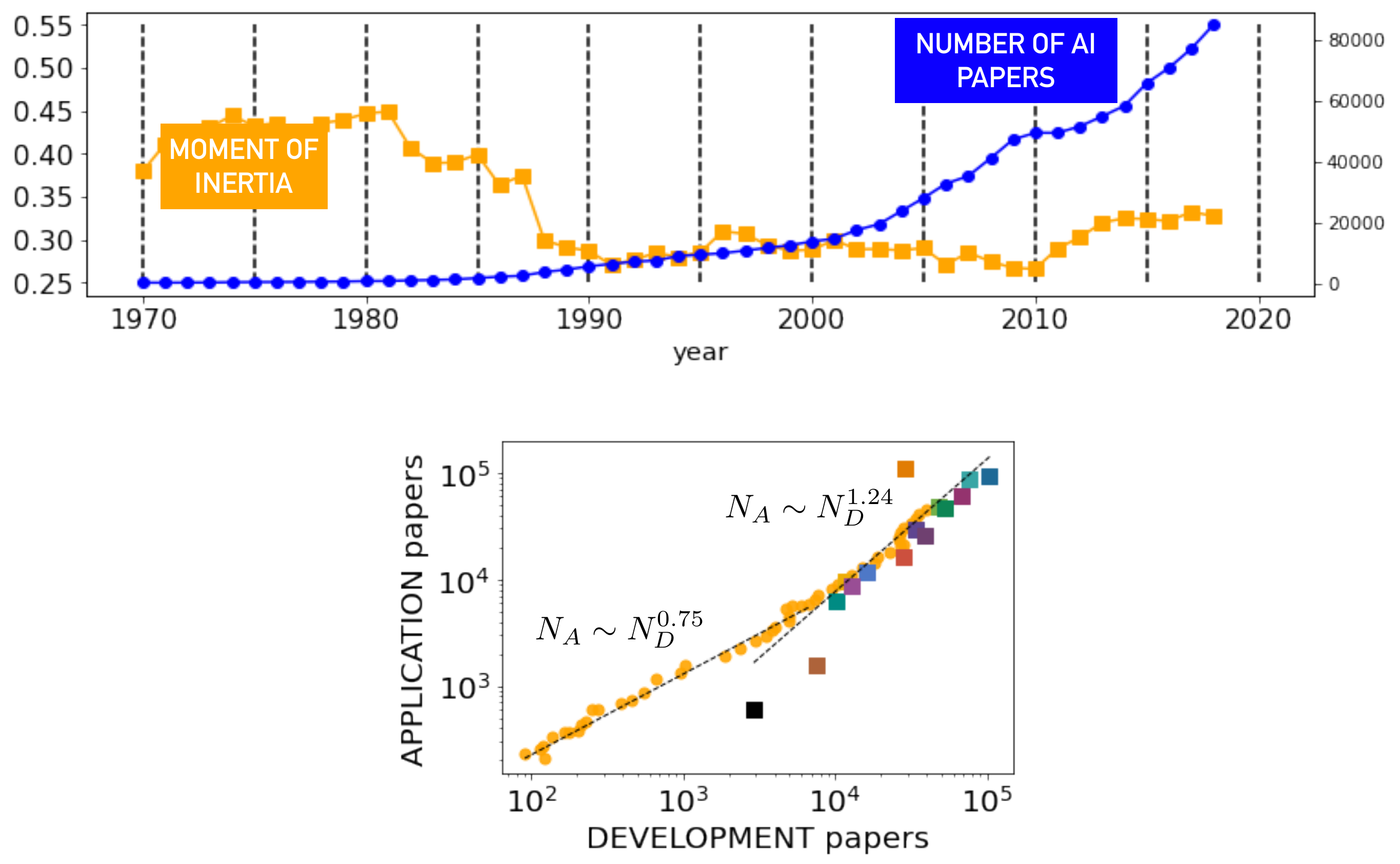}
\caption{AI development-vs-applications. Upper plot: Yellow line: moment of inertia. Blue line: total number of AI publications. Lower plot: number of application papers -vs- number of papers in native disciplines. Each yellow point represents one year. Each square represents temporal aggregation at the level of AI specialities.}
\label{fig3}
\end{figure}

For each year starting from 1970, we calculate the moment of inertia of the corpus. As outlined above, this indicator measures the dispersion of the corpus around the native disciplines. 
With this measure, the historical dynamics of AI appears as an oscillation between periods marked by forms of disciplinary dispersion followed by periods of disciplinary concentration.\\
Figure \ref{fig3} shows that before 1988, AI was present in numerous disciplines beyond computer science, mathematics and statistics (high moment of inertia). The so-called native disciplines were not the exclusive founders of AI whose origins appear to be much more interdisciplinary, with inputs from, among others, engineering, philosophy and psychology.
In 1988 a phase of concentration around native disciplines began, reaching a maximum in 2010 (low $m_I$). After 2010, the moment of inertia starts to increase again, indicating the gradual spreading of AI knowledge to other disciplinary domains, more distant from the native disciplines. Of note, the recent diffusion process started with a delay of around ten years after the take-off of scientific production in AI (around the year 2000).

Therefore, we observe cycles: a first phase of disciplinary diversity in the AI  ecosystem, then concentration (at the time of the AI winter and the emergence of the expert systems tradition) followed by a recent diffusion process (linked to the renewed interest in AI connected to DL applications). 

In the lower plot of Figure~ \ref{fig3} we analyze the relationship between the number of  papers in the native disciplines and the number of application papers. We do this analysis both at the level of years (yellow points) and at the level of specialities (coloured squares). The scaling shows the presence of two different regimes: when the number of native papers (in computer science, mathematics and statistics) is low ($<10,000$) we observe a sub--linear regime, in that applications grow more slowly than development of new concepts and tools. When the number of papers in native disciplines is high (and this trend is also confirmed by the aggregated values in terms of specialities) we are in a super--linear regime: the number of applications grows faster than the development production, i.e. each  paper in a native discipline gives rise to more than an application paper.

\begin{figure}[ht]
\centering
\includegraphics[width=1\linewidth]{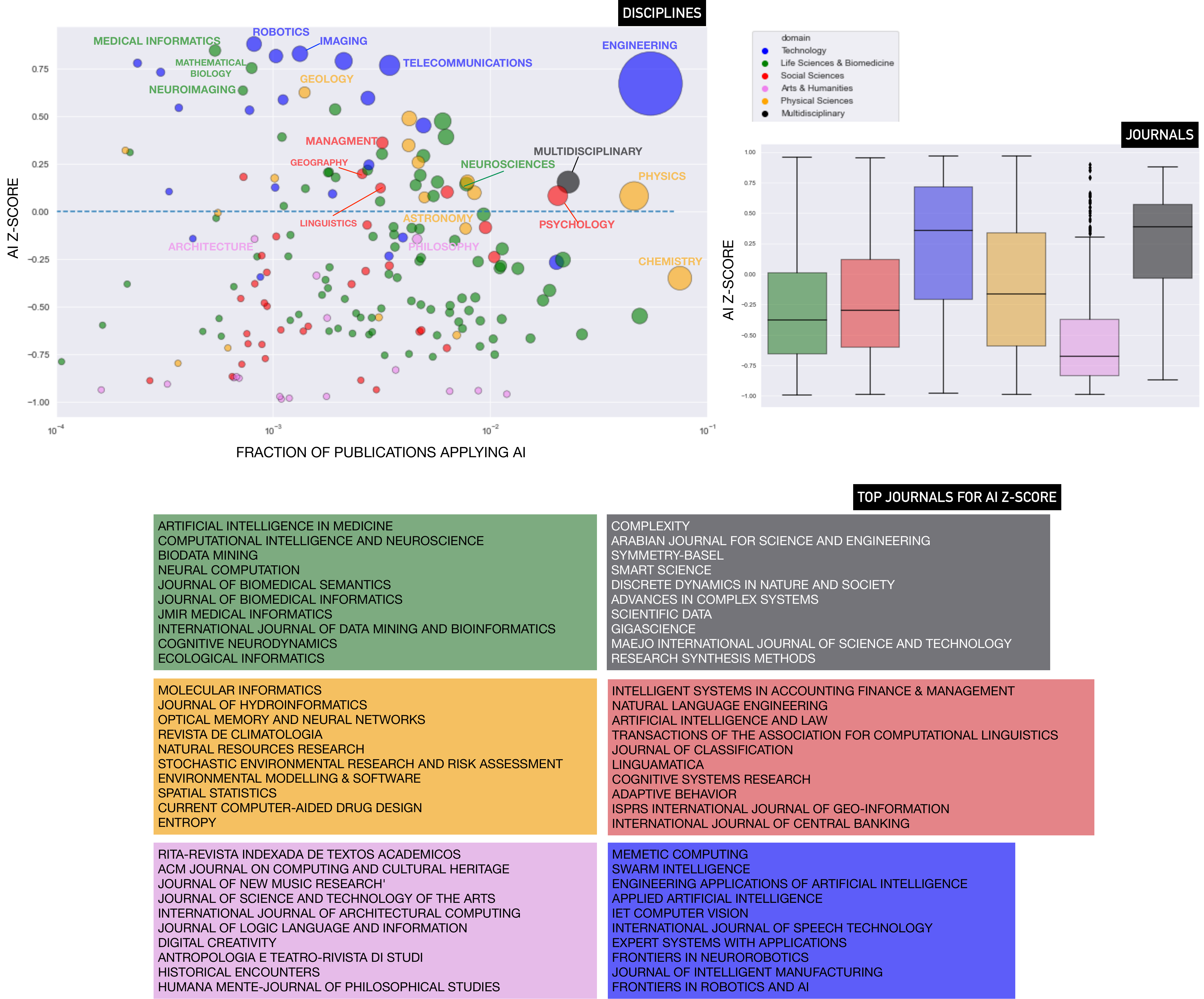}
\caption{AI application disciplinary landscape. Upper left plot: AI score -vs- fraction of AI papers for all disciplines. The size of the points is proportional to the total number of AI publications. Color indicates the WoS disciplinary category. Upper right plot: boxplot of the AI score of the journals in the six WoS disciplinary categories. Lower boxes: list of the top 10 journals publishing AI papers in each category.}
\label{fig4}
\end{figure}

After describing the aggregate scenario we explore the disciplinary composition of the AI applied ecosystem. As could be expected, Figure \ref{fig4} shows that technological disciplines (such as engineering, robotics, imaging) are the sectors in which AI is more largely over--represented. Some technical medical disciplines, like neuroimaging and medical informatics, are also intensely adopting AI methodologies. 
Our disciplinary AI score shows that the physical sciences are not always well positioned. For example, AI techniques are less prevalent in physics than in some social sciences fields such as (following WoS classification) management, geography or linguistics. Only the arts and humanities are consistently underrepresented. 

This pattern can be also described at the granularity scale of journals where we observe a dominance of AI in technology and multidisciplinary outlets. 
It is important here to keep in mind that having excluded conferences for which categorization is less fine, the physical sciences, life sciences, and social sciences are mostly at the same level. However in all categories, the journals that publish most AI-related papers are among those that specifically focus on computational methods. In the multidisciplinary journal landscape, especially noticeable is the presence of journals related to complex systems, which like AI can be seen as a technological platform, with multiple  contact points with AI techniques  (\cite{vigni2021complexity}).

\begin{figure}[ht]
\centering
\includegraphics[width=0.8\linewidth]{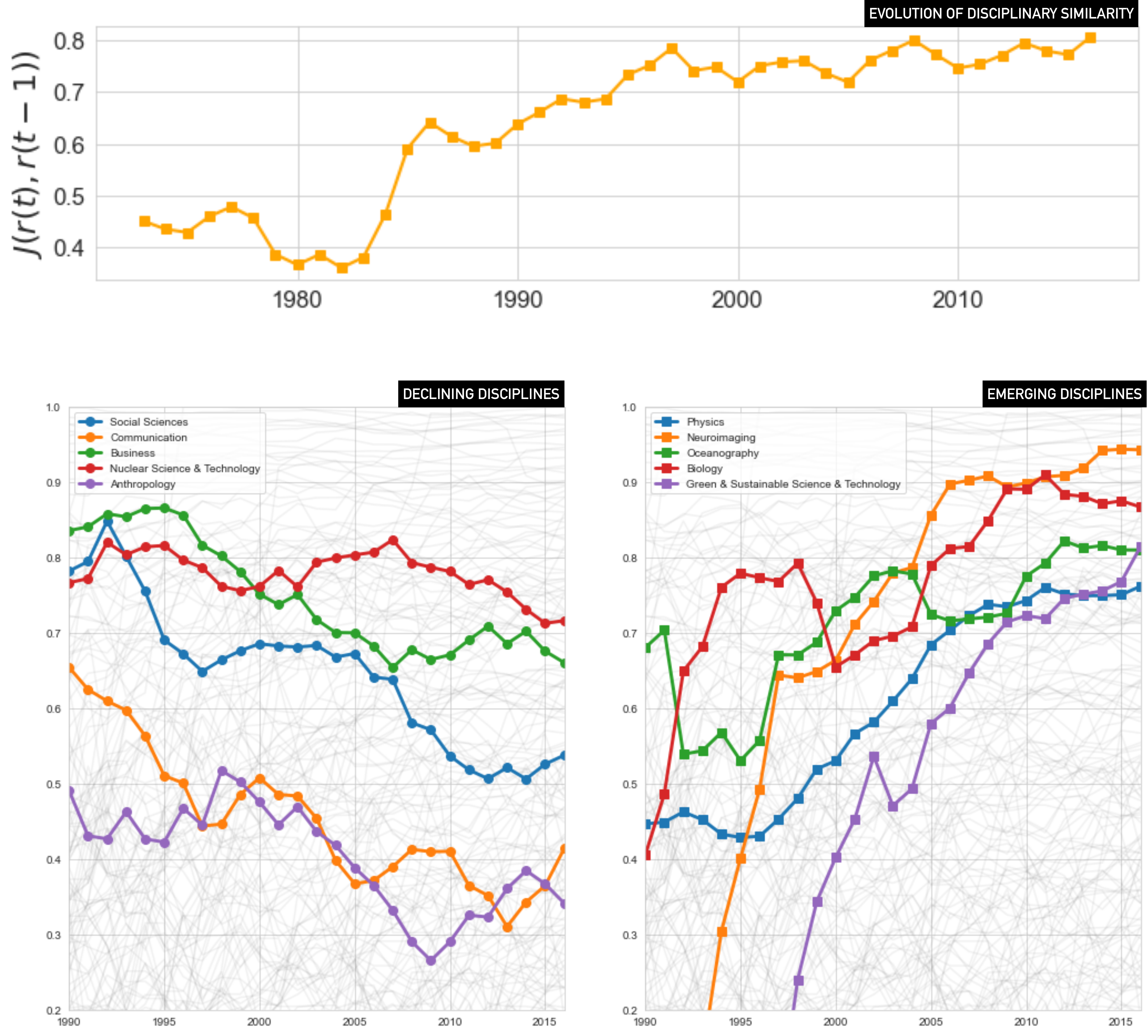}
\caption{AI application temporal disciplinary landscape. Upper plot: ranked Jaccard similarity between disciplinary ranks in two subsequent years.}
\label{fig4b}
\end{figure}

Concerning the diffusion process of AI we can see that the disciplinary ranking gets quite stable since the late 1990s (Figure \ref{fig4b}). To measure distance among rankings we use the ``ranked Jaccard similarity'' introduced in (\cite{gargiulo2016classical}).
In the lower plots of Figure \ref{fig4b} we can observe some prototypical trajectories of disciplines that experienced an important change in the ranking from the 1990s until today. Some disciplines (above all in social science) experienced a strong decrease being indeed strongly connected to decreasing AI specialities like symbolic AI and Expert Systems. Disciplines like physics and biology show a periodic growth (with a constant trend) in AI adoption while others, like neuroimaging and green \& sustainability technologies, display a sudden climbing of the ranking since their creation. 

\begin{figure}[ht]
\centering
\includegraphics[width=1\linewidth]{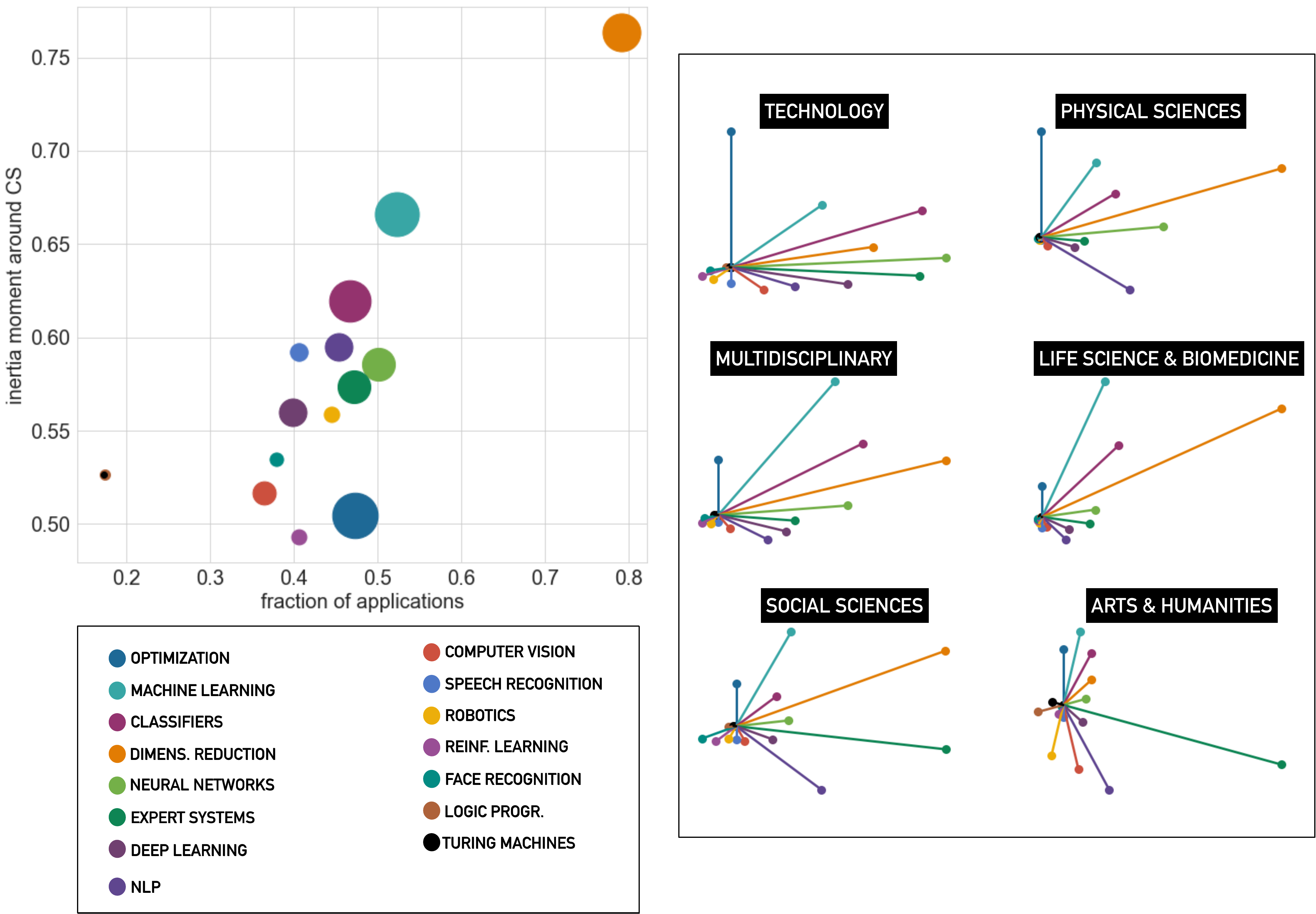}
\caption{The disciplinary landscape of AI specialities. Left plot: Moment of inertia around native disciplines -vs- fraction of applications, for all AI specialities. Right plot: share of all the different AI specialities to the main knowledge domains.}
\label{fig5}
\end{figure}

Examining specialities in more detail (Figure \ref{fig5}),. we observe that ``dimensionality reduction" is generally the most widespread in applications, in quantitative terms (fraction of applications with respect to native papers) and in terms of disciplinary distance from native disciplines. Instead, the performance of ``optimization'' (the more represented speciality in terms of total number of publications, as shown by the size of the point) is high in terms of fraction of applications, but very low in terms of the moment of inertia, namely it is largely applied in disciplines that are close to the native ones. 

The different knowledge domains have very diverse profiles in terms of the adoption of AI specialities. In the arts and humanities, applications are mostly  related to Expert Systems. The social sciences have strong interest in four AI specialities, namely machine learning, dimensionality reduction, expert systems and natural language processing (NLP). 
Physical sciences, as well as life sciences and multidisciplinary frameworks adopt dimensionality reduction, classifiers and machine learning. Technology disciplines have a much more uniform distribution on AI specialities. Optimization is relevant only for technology and, to a lesser extent, for the physical sciences. 

\subsection{Authors' collaborations in the AI landscape}
\label{authors}
The last part of this study analyzes the collaboration patterns driving the diffusion of AI. The basic question we address is whether the writing process of papers applying AI involves the direct collaboration of AI developers and experts in the application domains. 

\begin{figure}[ht]
\centering
\includegraphics[width=1\linewidth]{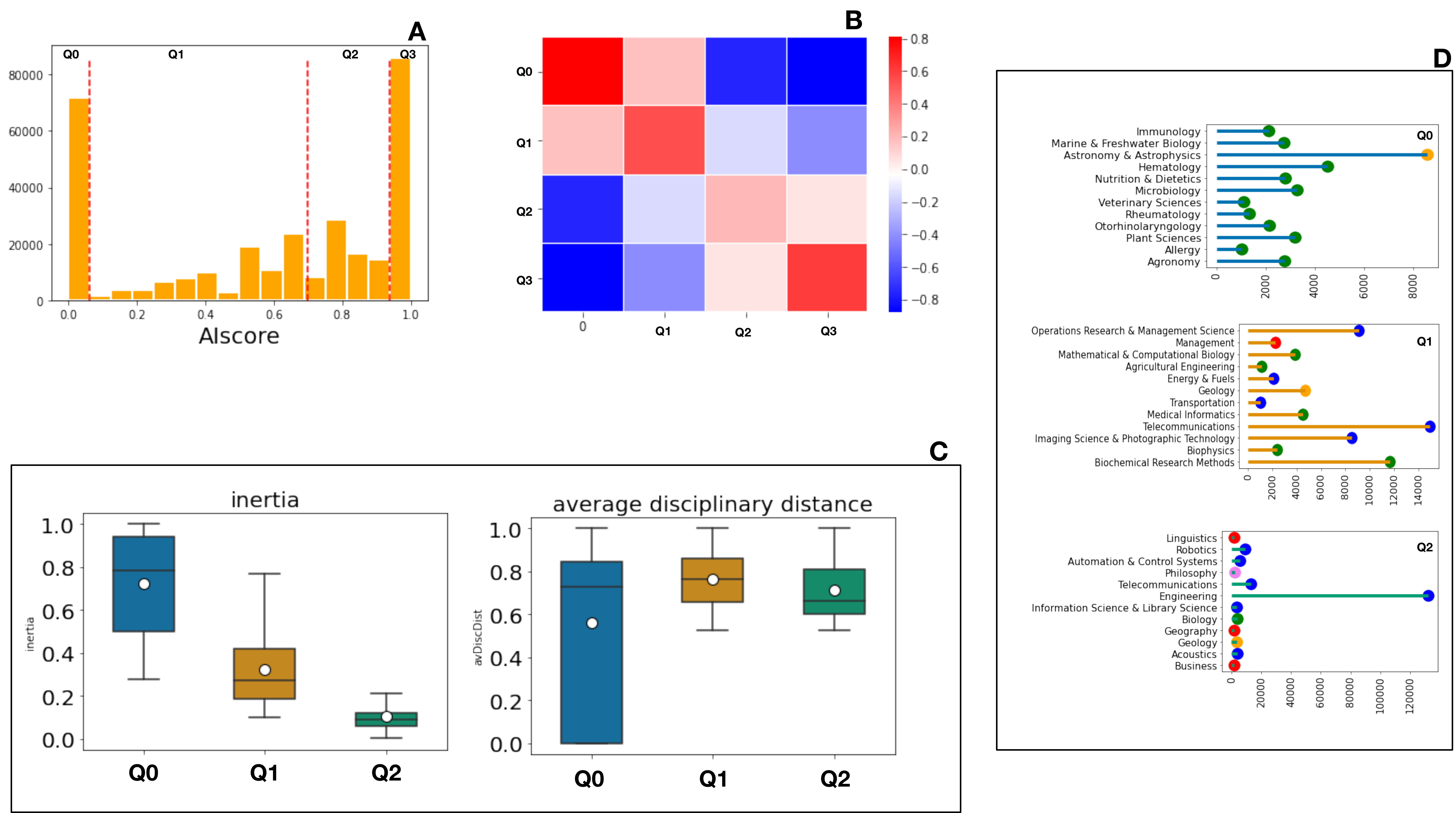}
\caption{AI collaborations. Plot A: Distribution of the authors' AI score, and identification of the quartiles. Plot B: Collaborations between groups of authors. The map refers to the comparison with the multinomial expectations for the links. Plot C: Box plots of the moment of inertia and of the inter-disciplinary distance for each group of authors. Plot D: Top 10 disciplines (sorted according to the disciplinary AI score in the group) in which the authors of each group specialize. The color of the points represents the knowledge domain associated to the discipline.}
\label{fig6}
\end{figure}

We assign to each author an AI score, $A_I$, as described above. The distribution of authors' AI scores shows two distinct peaks on the extremes, one around 0 (authors never publishing in AI journals) and the second around 1 (authors only publishing in AI journals). We divide the authors in groups according to the quartiles of their AI score: the first group (Q0) only contains authors publishing in out-disciplinary journals ($A_I=0$), the last one authors only publishing in AI journals (Q3, $A_I=1$), the two intermediate groups (Q1 and Q2) contain authors having respectively an AI-score lower and higher than 0.7 (Figure~\ref{fig6}A).

From the histograms in Figure~\ref{fig6}D we can also observe that authors in the first group (Q0) mostly publish in biomedical disciplines, but also astrophysics. Authors in Q1 are mostly specialized in interdisciplinary mathematical and technological applications (like mathematical biology, operation research, energy, telecommunications). Finally, Q2 includes mostly researchers in general engineering. 

Looking at Figure~\ref{fig6}C, authors in Q0 have the highest moment of inertia around native disciplines but at the same time have a low global level of interdisciplinarity. 
They interact with few disciplines, quite close to each other. Q1 includes most authors involved in interdisciplinary collaborations. Q2 authors publish in disciplines close to the native ones (low moment of inertia) but at a quite large distance from them.

Then we analyze the author collaboration network (ACN), the weighted graph structure where all authors present in the database are linked to all the other authors they collaborated with. Weight is given by the number of joint publications. Aggregating at the level of the four AI score groups of the authors, we calculate the fraction of collaborations between each class and all the others. To avoid the bias that might result from the different sizes of the groups, we compare these values to the expected multinomial distribution of these links (Figure~\ref{fig6}B). 
As we can observe from the figure, AI authors (those in the third and fourth quartiles of the AI-score distribution, Q2 and Q3) mostly collaborate among themselves. Authors in Q2, as observed before, are mostly involved in engineering disciplines and are therefore more connected to AI developers.
On the contrary, disciplinary experts (Q0 and Q1) have few collaborations with AI developers (Q3) and collaborate among them.

This highlights a clear separation among the scholars who do research \textit{on} AI and those who do research \textit{with} AI. 
Application of AI to other disciplines is not primarily driven by direct collaborations between authors with different backgrounds. However, there are some direct collaborations among the second and the third groups. To some extent, authors in these groups constitute a bridge between the theoretical development of AI and the disciplinary applications. 

\section{Discussion}
We have built a bottom-up cartography of AI, based on an analysis of the scientific production on AI, from 1970 until 2017. 

We started from an expert-based definition of AI trough the identification of all the terms that different actors involved in AI studies use to structure the vocabulary of this research field (collected as mentioned above, through an all-encompassing analysis of the web pages containing AI glossaries). 

We show the effects of what historians of science call the ``AI winters'', periods in which interest in AI declined. We also show that the AI scientific platform came out of these crises thanks to the innovation dynamics arising from the recombination of specialities: for example expert systems in the 1990s and DL around 2010 (DL was initially defined in 2006 and largely spread in 2012, with the ImageNet challenge).

The emergence of innovative specialities is partially due to the recombination of pre--existing building blocks from other domains but also from the dynamics of interdisciplinary participation. 

We therefore studied the historical spread of AI in the traditional disciplines. After the foundation of AI in a very interdisciplinary context,  we identified a first phase of disciplinary ``concentration'', during the ``AI winter'' during which the development of AI knowledge, previously distributed among several fields, condensed on the so-called native disciplines: computer science, mathematics and statistics. 
Only more recently, starting from 2010, we observe a new interdisciplinary phase of AI, with its gradual spreading to a larger number of fields where it is applied. 

Transfer of AI knowledge from development to application of AI is mediated by scientists in multiple fields of study, notably applied mathematical fields (like for example mathematical biology), geology, biophysics, and some engineering applied fields. 
A relevant role has been played in the last decade by multidisciplinary journals where applied AI papers from several fields were published.

The aim of this study was to reconstruct a bottom-up definition of AI, building a dynamic cartography of this domain from its published traces. Its deeper intent is, more generally, to be a precursor analysis on several directions of study connected to the more comprehensive ambition to understand the role of AI in the transformation of the scientific ecosystem.

For example the structure of AI specialities would require an in-depth qualitative study based on interviews of the actors involved in each of them, in order to investigate their overall perception of AI and their positioning in this quantitative landscape. A study of this type would be necessary to globally assess if AI can be really defined as a scientific platform (\cite{vigni2021complexity}) with a well--defined research program and objectives.

In this paper we mostly focused on the presence of AI terms in applied disciplines. 
We adopted the designation of ``native'' AI disciplines from the current literature (\cite{cockburn2018impact, bianchini2022artificial}) but our findings challenge it by showing how by its historical origin, AI was rather an interdisciplinary research area.
This interdisciplinary contribution was mostly evident in the historical practices commonly known as symbolic systems.
Later, different scientific fields have become, in turn, the central originating domains and applicators of AI knowledge, for example operational research which was for a long time one of the core actors of AI applications related to expert systems. 
A deeper historical analysis of the disciplines that developed AI, and the specialities of AI they focused on, would be worth of studying.

One way to investigate this question would be based on disciplinary case studies. A discipline can indeed be transformed by the introduction of a new set of knowledge, expanding its adjacent possible. Likewise, serendipitous interactions with external fields could spark new ideas. For example, neuroscience could be considered in principle as an originating domain of AI, notably concerning the development of neural network architectures, but the centrality of neuroscience journals in AI scientific production will need to be ascertained in detail.

This paper gives therefore important hints on how to navigate the AI scientific ecosystem in order to select potentially interesting case studies for subsequent analyses.


\section*{Acknowledgments}
This work is partially founded by the ANR project ScientIA. The PhD fellowship of S.F. is founded by the CNRS-MiTi program EpiAI. We thank M. F\"arber for  providing us with very useful information about the structure of the MAG database. 

\bibliography{sample}

\begin{thebibliography}{}

\bibitem[nat, 2019]{nature_2019}
 (2019).
\newblock The scientific events that shaped the decade.
\newblock {\em Nature}.

\bibitem[Aghion et~al., 2018]{aghion2018artificial}
Aghion, P., Jones, B.~F., and Jones, C.~I. (2018).
\newblock Artificial intelligence and economic growth.
\newblock In {\em The economics of artificial intelligence: An agenda}, pages
  237--282. University of Chicago Press.

\bibitem[Annoni et~al., 2018]{annoni2018artificial}
Annoni, A., Benczur, P., Bertoldi, P., Delipetrev, B., De~Prato, G., Feijoo,
  C., Macias, E.~F., Gutierrez, E.~G., Portela, M.~I., and Junklewitz, H.
  (2018).
\newblock Artificial intelligence: A european perspective.

\bibitem[Baum et~al., 2021]{baum2021artificial}
Baum, Z.~J., Yu, X., Ayala, P.~Y., Zhao, Y., Watkins, S.~P., and Zhou, Q.
  (2021).
\newblock Artificial intelligence in chemistry: current trends and future
  directions.
\newblock {\em Journal of Chemical Information and Modeling}, 61(7):3197--3212.

\bibitem[Bianchini et~al., 2020]{bianchini2020deep}
Bianchini, S., M{\"u}ller, M., and Pelletier, P. (2020).
\newblock Deep learning in science.
\newblock {\em arXiv preprint arXiv:2009.01575}.

\bibitem[Bianchini et~al., 2022]{bianchini2022artificial}
Bianchini, S., M{\"u}ller, M., and Pelletier, P. (2022).
\newblock Artificial intelligence in science: An emerging general method of
  invention.
\newblock {\em Research Policy}, 51(10):104604.

\bibitem[Blondel et~al., 2008]{blondel2008fast}
Blondel, V.~D., Guillaume, J.-L., Lambiotte, R., and Lefebvre, E. (2008).
\newblock Fast unfolding of communities in large networks.
\newblock {\em Journal of statistical mechanics: theory and experiment},
  2008(10):P10008.

\bibitem[Cockburn et~al., 2018]{cockburn2018impact}
Cockburn, I.~M., Henderson, R., and Stern, S. (2018).
\newblock The impact of artificial intelligence on innovation: An exploratory
  analysis.
\newblock In {\em The economics of artificial intelligence: An agenda}, pages
  115--146. University of Chicago Press.

\bibitem[F{\"a}rber, 2019]{farber2019microsoft}
F{\"a}rber, M. (2019).
\newblock The microsoft academic knowledge graph: A linked data source with 8
  billion triples of scholarly data.
\newblock In {\em International semantic web conference}, pages 113--129.
  Springer.

\bibitem[F{\"a}rber and Ao, 2022]{farber2022microsoft}
F{\"a}rber, M. and Ao, L. (2022).
\newblock The microsoft academic knowledge graph enhanced: Author name
  disambiguation, publication classification, and embeddings.
\newblock {\em Quantitative Science Studies}, 3(1):51--98.

\bibitem[Frank et~al., 2019]{frank2019evolution}
Frank, M.~R., Wang, D., Cebrian, M., and Rahwan, I. (2019).
\newblock The evolution of citation graphs in artificial intelligence research.
\newblock {\em Nature Machine Intelligence}, 1(2):79--85.

\bibitem[Gargiulo et~al., 2016]{gargiulo2016classical}
Gargiulo, F., Caen, A., Lambiotte, R., and Carletti, T. (2016).
\newblock The classical origin of modern mathematics.
\newblock {\em EPJ Data Science}, 5(1):26.

\bibitem[Kauffman, 2000]{kauffman2000investigations}
Kauffman, S.~A. (2000).
\newblock {\em Investigations}.
\newblock Oxford University Press.

\bibitem[King et~al., 2009]{king2009automation}
King, R.~D., Rowland, J., Oliver, S.~G., Young, M., Aubrey, W., Byrne, E.,
  Liakata, M., Markham, M., Pir, P., Soldatova, L.~N., et~al. (2009).
\newblock The automation of science.
\newblock {\em Science}, 324(5923):85--89.

\bibitem[Kitchin, 2014]{kitchin2014bigdata}
Kitchin, R. (2014).
\newblock Big data, new epistemologies and paradigm shifts.
\newblock {\em Big Data \& Society}, 1(1):1--12.

\bibitem[Li~Vigni, 2021]{vigni2021complexity}
Li~Vigni, F. (2021).
\newblock Complexity sciences: A scientific platform.
\newblock {\em Science \& Technology Studies}, 34(4):30--55.

\bibitem[McCarthy, 1981]{mccarthy1981epistemological}
McCarthy, J. (1981).
\newblock Epistemological problems of artificial intelligence.
\newblock In {\em Readings in artificial intelligence}, pages 459--465.
  Elsevier.

\bibitem[McCarthy, 2004]{mccarthy2004artificial}
McCarthy, J. (2004).
\newblock What is artificial intelligence.
\newblock {\em URL: http://www-formal. stanford. edu/jmc/whatisai. html}.

\bibitem[Monechi et~al., 2017]{monechi2017waves}
Monechi, B., Ruiz-Serrano, A., Tria, F., and Loreto, V. (2017).
\newblock Waves of novelties in the expansion into the adjacent possible.
\newblock {\em PloS one}, 12(6):e0179303.

\bibitem[OECD, 2019]{oecd}
OECD (2019).
\newblock {\em Artificial Intelligence in Society}.

\bibitem[Serrano et~al., 2009]{serrano2009extracting}
Serrano, M.~{\'A}., Bogun{\'a}, M., and Vespignani, A. (2009).
\newblock Extracting the multiscale backbone of complex weighted networks.
\newblock {\em Proceedings of the national academy of sciences},
  106(16):6483--6488.

\bibitem[Uzzi et~al., 2013]{uzzi2013atypical}
Uzzi, B., Mukherjee, S., Stringer, M., and Jones, B. (2013).
\newblock Atypical combinations and scientific impact.
\newblock {\em Science}, 342(6157):468--472.

\bibitem[WIPO, 2019]{wipo2019wipo}
WIPO (2019).
\newblock Wipo technology trends 2019: Artificial intelligence.
\newblock {\em Geneva: World Intellectual Property Organization}.

\end{thebibliography}

\end{document}